\documentclass[sigconf]{acmart}

\setcopyright{none}
\copyrightyear{}
\acmYear{}
\acmDOI{}
\acmConference[UKAIRS 2026]{UK AI Research Symposium}{November 24--25, 2026}{Edinburgh, UK}
\renewcommand\footnotetextcopyrightpermission[1]{}

\begin{document}
%% The "title" command has an optional parameter, allowing the author to define a "short title" to be used in page headers.

\title{Towards Transparent Computational Agents for Air Traffic Control:
Explainability, Interpretability and Trust}

%\title{Computational Agents for Air Traffic Control: Explainability, Interpretability and Trust}
%%
%% The "author" command and its associated commands are used to define the authors and their affiliations. Of note is the shared affiliation of the first two authors, and the "authornote" and "authornotemark" commands used to denote shared contribution to the research.
\author{Rob Procter}
\correspondingauthor
\authornotemark[1]
%\email{webmaster@marysville-ohio.com}
\affiliation{%
  \institution{University of Warwick \& Alan Turing Institute}
  \city{Coventry \& London}
  \country{UK}
  }
  \email{rob.procter@warwick.ac.uk}

\author{Mark Rouncefield}
\affiliation{%
  \institution{University of Siegen}
  \city{Siegen}
  \country{Germany}}
\email{m.rouncefield@lancaster.ac.uk}

\author{Richard Everson}
\affiliation{%
  \institution{Exeter University}
  \city{Exeter}
  \country{UK}}
\email{R.M.Everson@exeter.ac.uk}

\author{George De Ath}
\affiliation{%
  \institution{Exeter University}
  \city{Exeter}
  \country{UK}
  }
\email{g.de.ath@exeter.ac.uk}

\author{Nicholas Pepper}
\affiliation{%
  \institution{Alan Turing Institute}
  \city{London}
  \country{UK}
  }
\email{npepper@turing.ac.uk}

\author{Ed Henderson}
\affiliation{%
  \institution{Alan Turing Institute}
  \city{London}
  \country{UK}
  }
\email{ehenderson@turing.ac.uk}

\author{Elhassan Mohamed}
\affiliation{%
  \institution{Alan Turing Institute}
  \city{London}
  \country{UK}
  }
\email{emohamed@turing.ac.uk}

\author{Eseoghene Benjamin}
\affiliation{%
  \institution{Alan Turing Institute}
  \city{London}
  \country{UK}
  }
\email{ebenjamin@turing.ac.uk}

\author{Ben Carvell}
\affiliation{%
  \institution{NATS}
  \city{Fareham}
  \country{UK}
  }
\email{benjamin.carvell@nats.co.uk}

%% By default, the full list of authors will be used in the page headers. Often, this list is too long, and will overlap other information printed in the page headers. This command allows the author to define a more concise list of authors' names for this purpose.
\renewcommand{\shortauthors}{Procter et al.}

\begin{abstract}
We report preliminary findings from an ongoing investigation into transparency techniques for computational Air Traffic Control agents. Air Traffic Controller Officers (ATCOs) observed two types of agent in simulated air traffic controlling exercises. ATCOs found the chosen transparency techniques helpful to some extent: revealing how agents were making controlling decisions was an essential foundation for trust. However, the ATCOs’ trust depended on whether they could be made sense of in a live, operational context and the agents' actions where judged to be professionally competent.
\end{abstract}

%% The code below is generated by the tool at http://dl.acm.org/ccs.cfm. Please copy and paste the code instead of the example below.


%\ccsdesc[500]{Do Not Use This Code~Generate the Correct Terms for Your Paper}
%\ccsdesc[300]{Do Not Use This Code~Generate the Correct Terms for Your Paper}
%\ccsdesc{Do Not Use This Code~Generate the Correct Terms for Your Paper}
%\ccsdesc[100]{Do Not Use This Code~Generate the Correct Terms for Your Paper}

\maketitle
\vspace{-0.3cm}


\section{Introduction}
%The goal of Air Traffic Control (ATC) is the safe, orderly and expeditious movement of air traffic through controlled airspace.
%and Air Traffic Controllers (ATCOs) are responsible for making in real-time the decisions that ensure this goal is achieved. 
%Air Navigation Service Providers (ANSPs) are interested in whether computational agent-based systems can provide a solution to the challenge of managing growing volumes of air traffic. 
Project Bluebird is developing agent-based tools and exploring how they may assist ATCOs by proposing clearances\footnote{I.e., change flight level or heading} for aircraft. The safety-critical demands of air traffic control (ATC) make it important that ATCOs are able to assess an agent's trustworthiness: i.e., is the clearance suggested appropriate in the current context\cite{procter2025trustworthy}.

Transparency is a key requirement for users’ trust in AI systems \cite{valle2022transparency}. We report findings from studies examining how ATCOs evaluate agent transparency techniques for two agents: Hawk, a rule-based agent, and Falcon, an optimisation-based agent. 
%In the following section, we set out the conceptual framework that underpins our approach to transparency, explainability and interpretability. We then describe the transparency techniques that were implemented for evaluation, followed by our methodology and findings. The article concludes with a discussion of the implications of these findings for ATCO–agent teaming and directions for future research.

\section{Explainability and Interpretability} 
An agent possesses the property of transparency if it supports explainability and interpretability \cite{arrieta2020explainable}: i.e., \textit{how} it makes a decision and \textit{why} it makes a decision  \cite{ai2023artificial}.

We follow a socio-technical systems approach \cite{baxter2011socio} to understanding how to operationalise and evaluate these properties. In particular, we are interested in the use ATCOs, as members of a specific community of practice \cite{wenger2000communities}, make of \textit{members’ methods} -- shared understandings people use to achieve the mundane orderliness of their everyday activities -- to make sense of agent behaviour \cite{garfinkel2023studies}.

\begin{figure}[h]
  \centering
 \includegraphics[width=\linewidth]{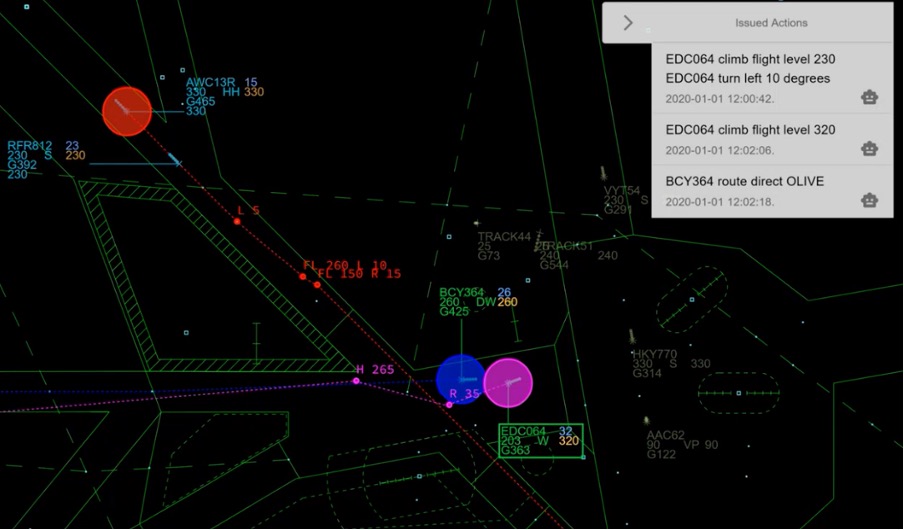}
  \caption{Falcon display showing relevant aircraft (orange, blue, green circles) and agent plan (magenta lines) for subject aircraft (green bounded box) and time ordered, scrolling list of current and recent clearances (top right).}
  \label{Figure 1}
\end{figure}

\vspace{-0.65cm}
\section{Visualising Relevant Aircraft and Plans}
Following discussions with ATCOs, \textit{relevant aircraft} and \textit{agent plans} were chosen as the explainability and interpretability techniques to be evaluated. 

In ATC, relevant aircraft is the subset of traffic that materially constrains a selected (subject) aircraft’s safe and efficient plan\footnote{I.e., could lead to a loss of separation. A minimum separation of 5 nautical miles or 1000ft is required to be maintained at all times}. It was chosen because it invites the observer to attend to what is driving the agent’s plans for the subject aircraft. 

Plans were chosen as the interpretability technique because they enable the observer to assess the agent's actions in context. Both these techniques are ATCO- rather than agent-centric, reflecting the methods that are constitutive of ATCOs' working practices \cite{garfinkel2023studies}.

For Hawk, any aircraft posing a risk to the subject aircraft of a loss of separation is identified as relevant. Hawk does not plan beyond the next clearance, however, ATCOs can infer from its history of clearances whether Hawk’s behaviour is consistent with having a plan \cite{carvell2026human}. 

Falcon optimises plans simultaneously for all aircraft. First it attempts to find the most efficient plan for an aircraft entering the sector and relevant aircraft are identified. Additional relevant aircraft are identified transitively with respect to these. Finally, subject and all relevant aircraft are jointly optimised \cite{mohamed2026towards} and these plans are be displayed (Figure \ref{Figure 1}). 

For both agents, a time ordered, scrolling list of agent clearances was displayed in the top right corner. 
%In the case of Falcon, a display of the agent’s plan for controlling the subject and all relevant aircraft for their entire transit through the sector could also be toggled on and off.

\section{Methodology}
%We have been evaluating how agent transparency
%, implemented through visualisation of relevant aircraft and agent plans, influences ATCOs’ trust: are the agents taking into account the right aircraft (explainability); given relevant aircraft and agent plans, can ATCOs make sense of the agents’ behaviour (interpretability); and how does this influence their trust in the agent?

In a series of formative evaluation sessions, ATCOs observed Hawk and Falcon controlling a simulated sector equipped with the transparency techniques described above. A 'think-aloud' methodology was followed, with ATCOs 
%Before the sessions began, the ATCOs were shown runs of each agent controlling and the underlying concepts of rule-based and optimising agents were explained
encouraged to comment on how well the techniques enabled them to understand agent behaviour \cite{monk1993improving}. %Agent developers were present and were able to respond to questions posed by the ATCOs.

\section{Findings}
Analysis centred around three themes: explainability (do agents identify the right relevant aircraft), interpretability (can agent plans be made sense of given the relevant aircraft), and the relationship between transparency and trust in agent performance. 
%We then examine two further themes that emerged from the data: modes of ATCO-agent interaction and teaming, and agent transparency requirements beyond of the live controlling environment.

\subsection{Explainability: is the agent identifying the right aircraft?} 
ATCOs commented on whether the agents identified relevant aircraft correctly and, if not, what kind of mistakes did the agents make, and where ATCOs able to diagnose the reasons. In this extract, the ATCOs’ question Hawk’s identification of relevant aircraft:

\begin{quote}
''It’s this longitudinal catch up situation... the catch up speed is 12 knots, which is basically nothing... TomJet was never really relevant traffic.''
\end{quote}

The ATCOs' assessed Hawk was overly cautious about the risk that a faster aircraft will catch it up a slower one (TomJet), because the rate of catch-up was too slow. In such ways, visualising relevant aircraft enabled ATCOs to diagnose the agents' controlling behaviour.

\subsection{Interpretability: is the agent issuing the right clearances?}
%ATCOs' attention would frequently shift from explainability to interpretability. 
%The ATCOs interpreted agent behaviour by comparing  agent clearances with what they would have done. In so doing, ATCOs often made use of members' methods, i.e., their shared understandings of what distinguishes good from poor controlling: 
%from relevant aircraft to the quality of the agent’s controlling, i.e., 
%ATCOs' comments frequently focused on the quality of the agent’s controlling when making sense of its behaviour:

Hawk's relative simplicity meant its behaviour was generally easy to interpret. In this extract, however, the ATCOs were baffled by its failure to put AKK8 back on its own navigation, i.e., what a competent ATCO would have done:

\begin{quote}
%''Relevant traffic is definitely helping to explain behaviours. It’s just that [its] behaviours aren’t necessarily always correct.''
''AKK8 could be back on its own navigation... we don’t understand why it’s still being vectored... it’s more over controlling.''
\end{quote}

The ATCOs acknowledged Falcon's benefit in presenting complete plans for each aircraft, but its complexity made its behaviour more difficult to interpret:

\begin{quote}
''Hawk is showing you immediately what’s relevant to these planes. Falcon is showing me not only what’s relevant, but what I intend to do about it with all of the other flights... it is a lot to interpret... there’s pluses and minuses to both.''
\end{quote}

\subsection{Transparency and trust}
The study reveals how explainability and interpretability contribute to agent transparency. Relevant aircraft shows promise in satisfying explainability and relevant aircraft and plans together helped the ATCOs to interpret agent behaviour. Transparency doesn't guarantee trust, however, it enables deciding if the agent's behaviour can be trusted and this trust is calibrated through experience:
%making sense of the agent’s actions is more complicated. The remark ''we don’t understand'' in the previous extract speaks to ATCOs' recurring difficulties in trusting the agent’s behaviour. 

%In the following extract the challenge of making sense of Hawk’s actions is compared to Falcon’s:

\begin{quote}
''When it [the agent] fails to implement an early plan for a pair of aircraft, you become a lot more negative... it's also doing it this time. So, you build up a sort of increasing amount of negativity towards it.''
\end{quote}

An agent’s behaviour can be transparent and safe, yet not be trusted if its controlling does not align with the ATCOs' standards.

%''Hawk is a very straightforward agent to understand because it has been programmed with the rules of the air traffic game... 
%It’s not making what appear to be illogical decisions such as Falcon was demonstrating... 
%Falcon is trying to display trajectories and proposed clearances for all of the aircraft... 
%But air traffic is quite a complicated task, 
%so it becomes clunky to try to understand it all the time... 
%It may be that Falcon would become clearer with greater use... 
%if I were to watch it more... 
%at the moment, I think it’s very difficult for a human observer to use it.”
%\end{quote}

\section{Discussion}
Our findings suggest agent explainability can be satisfied by visualising relevant aircraft, while agent interpretability resides in the relationship between what relevant aircraft reveal about the agent’s reasoning and the clearances it issues.

ATCOs made sense of an agent's behaviour in terms of whether relevant aircraft were correctly identified and if the plans were appropriate give those relevant aircraft.
%and were able to diagnose reasons for poor agent control. 
Importantly, the ATCOs frequently judged the agents to fall short against the standards of good controlling and their trust in the agents was calibrated on this basis.

ATCOs found Hawk's behaviour easier to interpret than that Falcon's. More work is needed to devise visualisation techniques that are better matched to agent complexity \cite{mohamed2026towards}.

\section{Conclusions and Future Work}
ATCOs experienced agent transparency as a situated, ATCO-centred, achievement. Members' methods featured prominently in the ways ATCOs queried choices of relevant aircraft and the quality of agent plans. ATCOs made sense of what they observed by moment-by-moment assessment of the state of the sector: which aircraft they judged to be relevant and what would be a good plan. In these ways, ATCOs calibrated their trust in the agent: their confidence in the agent’s reasoning (relevant aircraft), together with scrutiny of the quality of the agent’s behaviour (plans). 

People's trust in the tools they use is a volatile relationship; it is calibrated and sustained through practices distributed across the tool life cycle: development and testing, the training people receive in their use and their lived experience of working with them. 

%Members’ methods are a central part of ATCOs’ community of practice \cite{Proctertrust}. If agents are to be embedded successfully in ATC, and in High Reliability Organisations in general, then more research is needed so that members’ methods can evolve in ways that ensure human-agent teaming is safe and effective. 
The members’ methods ATCOs drew on in this study to calibrate the trustworthiness of the agents are a central part of ATCOs’ community of practice \cite{Proctertrust}. If agents are to be embedded successfully in ATC, and in High Reliability Organisations in general, then more research is needed so that members’ methods can evolve in ways that ensure human-agent teaming is safe and effective. 

\begin{acks}
Project Bluebird is a Prosperity Partnership between NATS, the University of Exeter and the Alan Turing Institute, funded by EPSRC (Grant Number EP/V056522/1).
\end{acks}

%\section{Figures}
%The ``\verb|figure|'' environment should be used for figures. One or more images can be placed within a figure. If your figure contains third-party material, you must clearly identify it as such, as shown in the example below.
%\begin{figure}[h]
%\centering
%\includegraphics[width=\linewidth]{sample-franklin}
%\caption{1907 Franklin Model D roadster. Photograph by Harris \& Ewing, Inc. [Public domain], via Wikimedia Commons. (\url{https://goo.gl/VLCRBB}).}
%\Description{A woman and a girl in white dresses sit in an open car.}
%\end{figure}

%Your figures should contain a caption which describes the figure to the reader.

%Figure captions are placed {\itshape below} the figure.

%Every figure should also have a figure description unless it is purely decorative. These descriptions convey what’s in the image to someone who cannot see it. They are also used by search engine crawlers for indexing images, and when images cannot be loaded.

%% The next two lines define the bibliography style to be used, and the bibliography file.
\bibliographystyle{ACM-Reference-Format}
\bibliography{sample-base}
\end{document}